\documentclass[aps,twocolumn,amsmath,amssymb,floatfix,prx,reprint,showpacs,footinbib,superscriptaddress,longbibliography]{revtex4-1}
\usepackage{epsfig,graphicx}
\usepackage{setspace}
\usepackage[english]{babel}
\usepackage{amsfonts}
\usepackage{amsmath}
\usepackage{latexsym}
\usepackage{graphics,bm}
\usepackage{natbib}
\usepackage{dcolumn}
\usepackage{bm}
\usepackage{rotating}
\usepackage{amsmath}
\usepackage{braket}
\usepackage{epstopdf}
\usepackage{color}

\usepackage{soul,xcolor}
\setstcolor{red}
\usepackage[T1]{fontenc}
\usepackage[applemac]{inputenc}
\usepackage{lmodern}
\usepackage{ae}
\usepackage{units}
\usepackage{color}
\usepackage{url}
\usepackage[colorlinks]{hyperref}
\hypersetup{%
        plainpages=true,
        breaklinks=true,
        hypertexnames=false,
        pageanchor=true,
        colorlinks=true,
        linkcolor={blue},
        citecolor={red},
        urlcolor={blue},
        anchorcolor={black}
      }
\newcommand{\be}{\begin{equation}}
\newcommand{\ee}{\end{equation}}
\newcommand{\ba}{\begin{array}{lll}}
\newcommand{\ea}{\end{array}}

\newcommand{\X}{{\,_{\!}}}

\begin{document}
\title{Non equilibrium photonic transport and phase transition in an array of optical cavities}
\author{Kamanasish Debnath}
\email[e-mail:]{kamanasish.debnath@gmail.com}
\affiliation{Institute of Physics, Ecole Polytechnique F\'ed\'erale de Lausanne (EPFL), CH-1015 Lausanne, Switzerland}
\author{Eduardo Mascarenhas}
\email[e-mail:]{eduardo.mascarenhas@epfl.ch}
\affiliation{Institute of Physics, Ecole Polytechnique F\'ed\'erale de Lausanne (EPFL), CH-1015 Lausanne, Switzerland}
\author{Vincenzo Savona}
\email[e-mail:]{vincenzo.savona@epfl.ch}
\affiliation{Institute of Physics, Ecole Polytechnique F\'ed\'erale de Lausanne (EPFL), CH-1015 Lausanne, Switzerland}
\date{\today}
\begin{abstract}
We characterize photonic transport in a boundary driven array of nonlinear optical cavities. We find that the output field suddenly drops when the chain length is increased beyond a threshold. After this threshold a highly chaotic and unstable regime emerges, which marks the onset of a super-diffusive photonic transport. We show the scaling of the threshold with pump intensity and nonlinearity. Finally,
we address the competition of disorder and nonlinearity presenting a diffusive-insulator phase transition.
\end{abstract}
\maketitle
\section{Introduction}\label{I}
Arrays of optical nonlinear cavities have been a field of extensive research \cite{Noh2017,Greiner2002,Schir2016,PhysRevA.90.023827,PhysRevA.92.063817,PhysRevLett.117.213603,PhysRevLett.98.180601,PhysRevA.78.062338,PhysRevLett.101.246809,PhysRevX.4.031043,PhysRevLett.116.235302,Brantut1069,LPOR:LPOR200810046,Tomadin:10,Jin2013,PhysRevA.90.023827,PhysRevA.91.033823,PhysRevA.76.031805}
since the low temperature analogy to condensed matter was established, suggesting the possibility of simulating the Mott-Superfluid phase transition \cite{Plenio2006, Greentree2006}. However, due to the natural driven dissipative character of optical systems, the acclaimed equilibrium Mott-Superfluid transition is generally absent, and can be approached only by means of a highly engineered dissipative environment \cite{Lebreuilly2017,Biella2017}.
The field has progressed along two main lines. The first is the nonequilibrium synthesis and stabilization of strongly correlated phases such as Mott \cite{Biella2017,Lebreuilly2017} and solid phases \cite{Jin2013} in analogy to the ground state physics. The second is the identification and characterization of genuine nonequilibrium phases and transitions such as the emergence of the gas-liquid bistability \cite{Biondi2016} and spontaneous symmetry breaking \cite{Savona2017}. Most of this activity has been restricted to theoretical work of homogeneous systems.

Very recently, an experiment was carried out~\cite{PhysRevX.7.011016}, reporting on indications of a driven-dissipative quantum phase transition and existence of bistable phases in a boundary driven array of circuit QED resonators. This pioneering work opens up the possibility to experimentally address nonequilibrium transport of interacting photons that is currently at its infancy even on theoretical grounds~\cite{PhysRevA.91.053815,PhysRevA.94.013809}. In this context, the present work aims at analyzing how the photonic current scales with the system size in presence of optical nonlinearity and local disorder.

We consider a boundary driven Bose-Hubbard chain and find that in general the system is either a (generalized) diffusive conductor, or an insulator. More precisely,
we show that, even though small systems may appear as stable ballistic conductors, there exists a ``critical" chain length $N_c$, beyond which the system becomes chaotic. This chaotic instability marks the onset of super-diffusive transport associated to power law scaling of the currents.
We also show that there are threshold values of disorder above which the scaling of the current becomes exponentially vanishing on the system size, thus marking the insulating phase.
We characterize the phase diagram and discuss experimental proposals to observe similar effects in cQED hybrid systems for future experiments.

The structure of the paper is as follows. We describe our model and setup in Sec.~\ref{II}, followed by our results in Sec.~\ref{III}, where we characterize the transport behaviour as a function of pump amplitude and nonlinearity, followed by the scaling of threshold chain lengths as a function of the system parameters. In Sec.~\ref{IV}, we perform a verification of the origin of chaos from the perspective of stability analysis and present results on how the system transits into chaotic behavior. In Sec.~\ref{V}, we show that the model undergoes a diffusive-insulator phase transition at sufficiently high disorder and present the phase portrait. We summarize in Sec.~\ref{VI} and present an outlook for future work.

\begin{figure}
\centering
\includegraphics[width=0.45\textwidth]{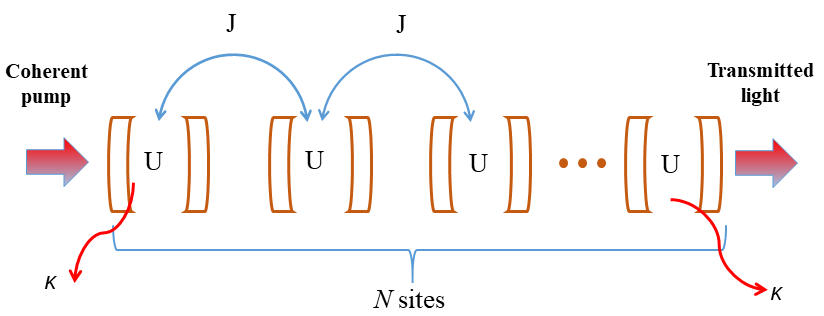}
\caption{The schematic representation of the boundary driven and dissipative non linear arrays of optical cavities.}
\label{schematic}
\end{figure}

\section{The model}\label{II}
We study photonic transport in a one dimensional Bose-Hubbard chain of $N$ coupled optical cavities with resonant frequency $\omega_c$, in presence of a coherent pump of frequency $\omega_L$ injecting photons at the first site and dissipation $\kappa$ only at the chain boundaries. The coupling between neighboring cavities is characterized by a hopping coefficient $J$, which competes with the local on-site Kerr nonlinearity $U$ in determining the localization of the photons. The Hamiltonian of such a system, in the rotating frame of the cavities, takes the form-
\be
H= \sum_i-\delta \hat{a}_i^{\dagger}\hat{a}_i^\X + U\hat{a}_i^{\dagger}\hat{a}_i^\X \hat{a}_i^{\dagger}\hat{a}_i^\X + J(\hat{a}_{i+1}^{\dagger}\hat{a}_i^\X + h.c) + p(\hat{a}_1^\X + \hat{a}_{1}^{\dagger})
\label{E1}
\ee
where $p$ is the amplitude of the driving field acting on the first site and $\delta= \omega_L - \omega_c$ is the detuning of the driving field. $a_i^\X$ and $a_i^{\dagger}$ are the bosonic annihilation and creation operators respectively at $i^{th}$ site, obeying the commutation relations $[a_i^\X,a _j^{\dagger}]=\delta_{i,j}$ and $[a_i^\X (a_i^{\dagger}),a_j^\X(a_j^{\dagger})]=0$. A pictorial representation of the model has been shown in Fig.~\ref{schematic}. Solving Eq.~(\ref{E1}) exactly in the quantum limit is a demanding task even in terms of best available computational resources. We consider mean field approximation such that $a_i= \alpha_i + \delta a_i$, where $\alpha_i$ is the average value of the photonic field of the $i^{th}$ site and $\delta a_i$, their corresponding quantum fluctuation. Under the classical approximation, $\delta a_i=0$. This is valid in the regime where the quantum fluctuations are negligible compared to the average value of the fields \textit{i.e.} $\langle\delta a_i \delta a_j\rangle << \alpha_i\alpha_j$. The regimes investigated in this article that produces non-trivial physics is inaccessible with exact quantum mechanical formulation and hence the validity of our approximation still remains open. However, mean field approximation was previously shown to be in perfect agreement in the context of a driven-dissipative quantum phase transition for a similar model~\cite{PhysRevX.7.011016}. The time derivative of the classical field equations at different sites can be written as-

\begin{figure}
\centering
\includegraphics[width=0.5\textwidth]{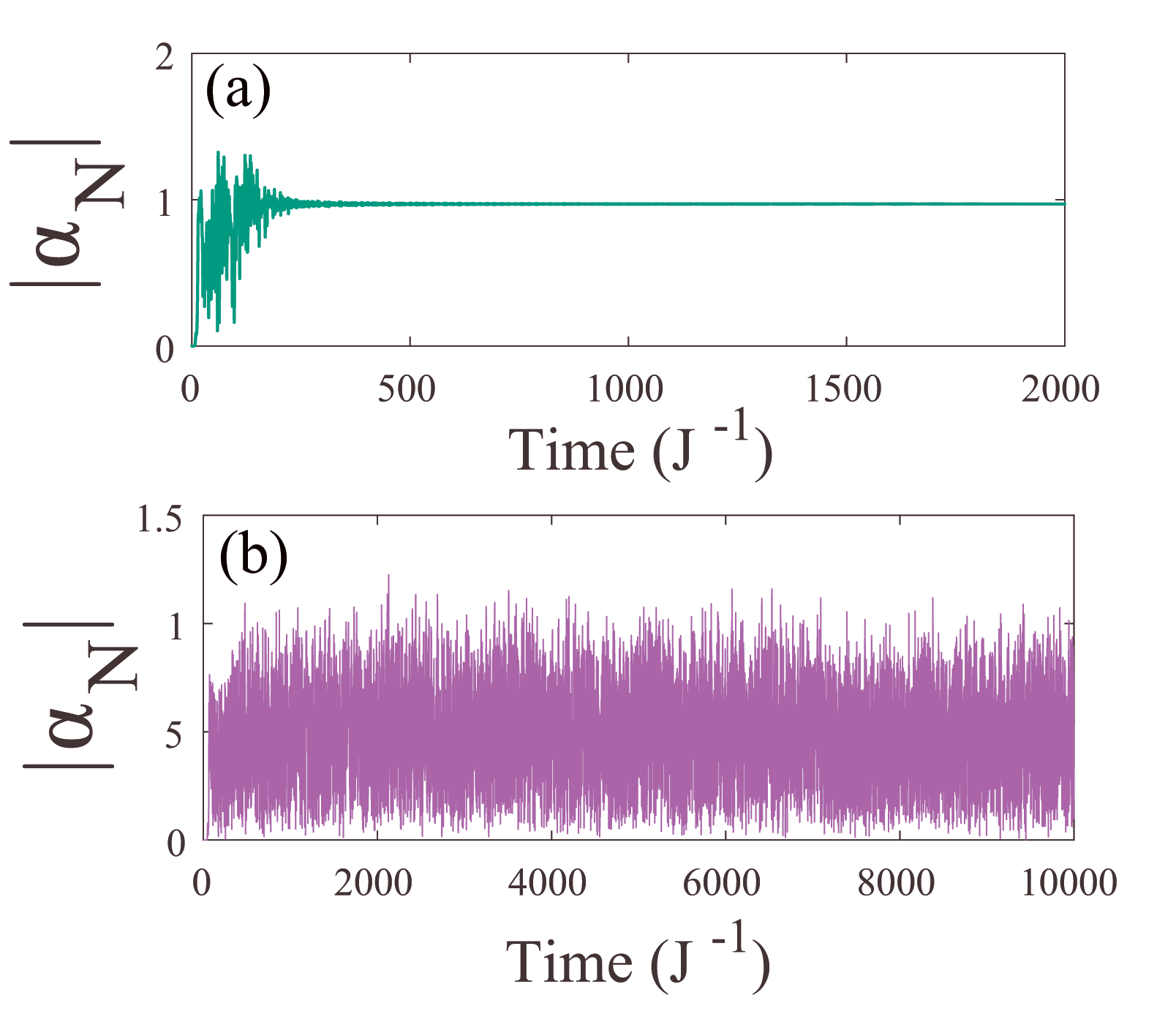}
\caption{The time dynamics of the output field intensity $\langle|\alpha_N|\rangle$ showing the stable regime for (a) $N= 20$ and the unstable regime for (b) $N= 100$.  Other parameters: $U/J=1.0$, $p/J=10$, $\delta= 0$ and $\kappa=J$.}
\label{F0}
\end{figure}

\be
\dot{\alpha}_1 = -\Big(\frac{\kappa}{2} + i\delta\Big)\alpha_1 - iJ\alpha_2 -2iU\mid \alpha_1 \mid^2 \alpha_1 -ip
\label{E2}
\ee
\be
\dot{\alpha}_{i} = -i\delta \alpha_1 -iJ\alpha_{i+1} - iJ\alpha_{i-1} -2iU\mid\alpha_i\mid^2 \alpha_i
\label{E3}
\ee
\be
\dot{\alpha}_N = -\Big(\frac{\kappa}{2}+ i\delta\Big)\alpha_N  - iJ\alpha_{N-1} -2iU\mid\alpha_N\mid^2\alpha_N
\label{E4}
\ee
where Eq.~(\ref{E3}) represents the equation for $i^{th}$ field with $i= 2, 3, .... N-1$, while Eqs.~(\ref{E2}) and (\ref{E4}) govern the field of the first and last boundary sites respectively. In this model, we further assume that dissipation at rate $\kappa$ only acts at the boundary sites.

\begin{figure}
\centering
\includegraphics[width=0.5\textwidth]{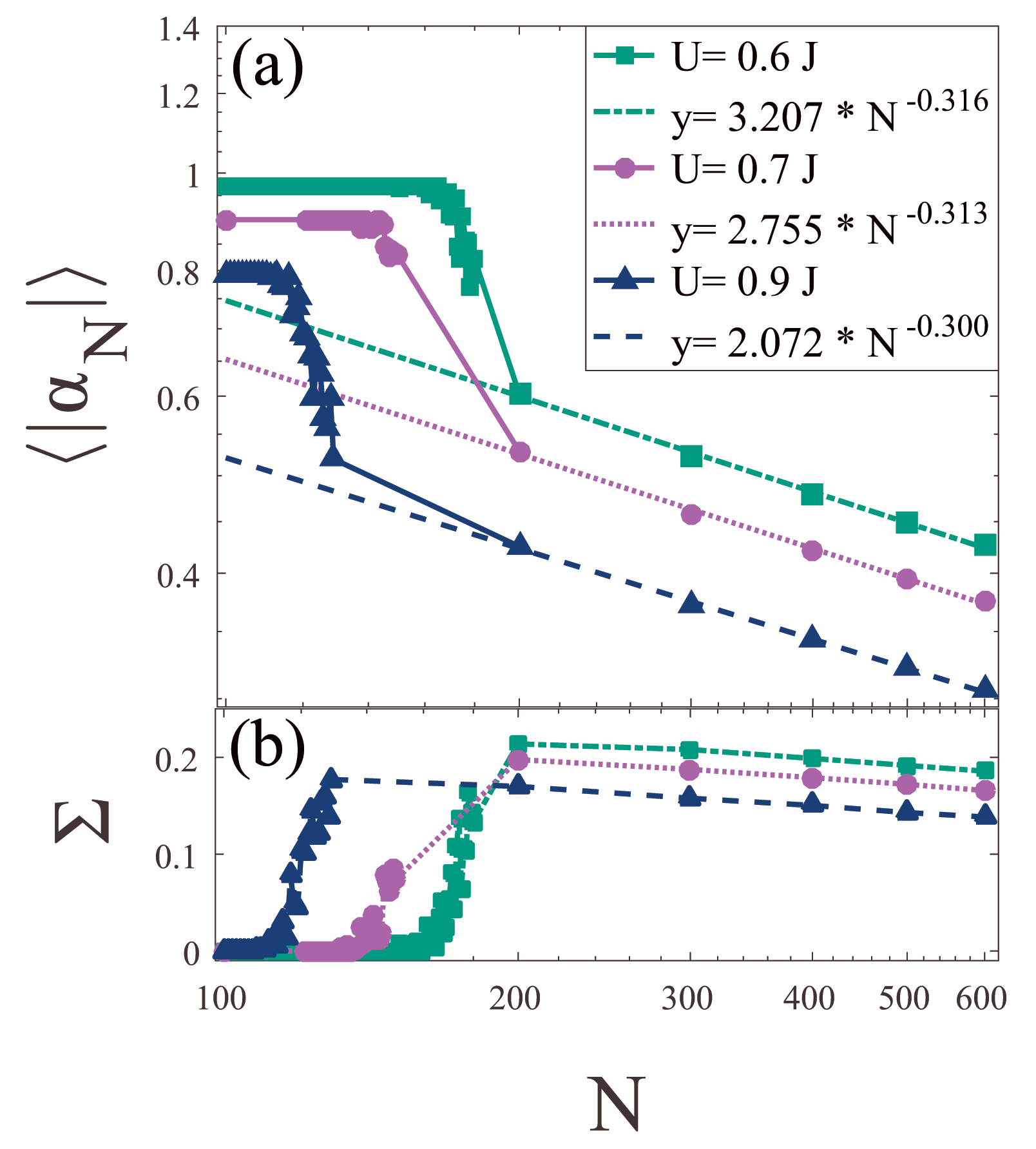}
\caption{(a) Average transmitted field intensity $\langle|\alpha_N|\rangle$ as a function of chain length $N$ for different values of $U$ at fixed pump $p=5J$. (b) The corresponding variance $\Sigma \equiv \sqrt{\langle|\alpha_N^2|\rangle - \langle|\alpha_N|\rangle^2}$ of the field intensity. Markers denote the data points and their corresponding fits are shown with dotted lines.}
\label{F1}
\end{figure}

\begin{figure}
\centering
\includegraphics[width=0.47\textwidth]{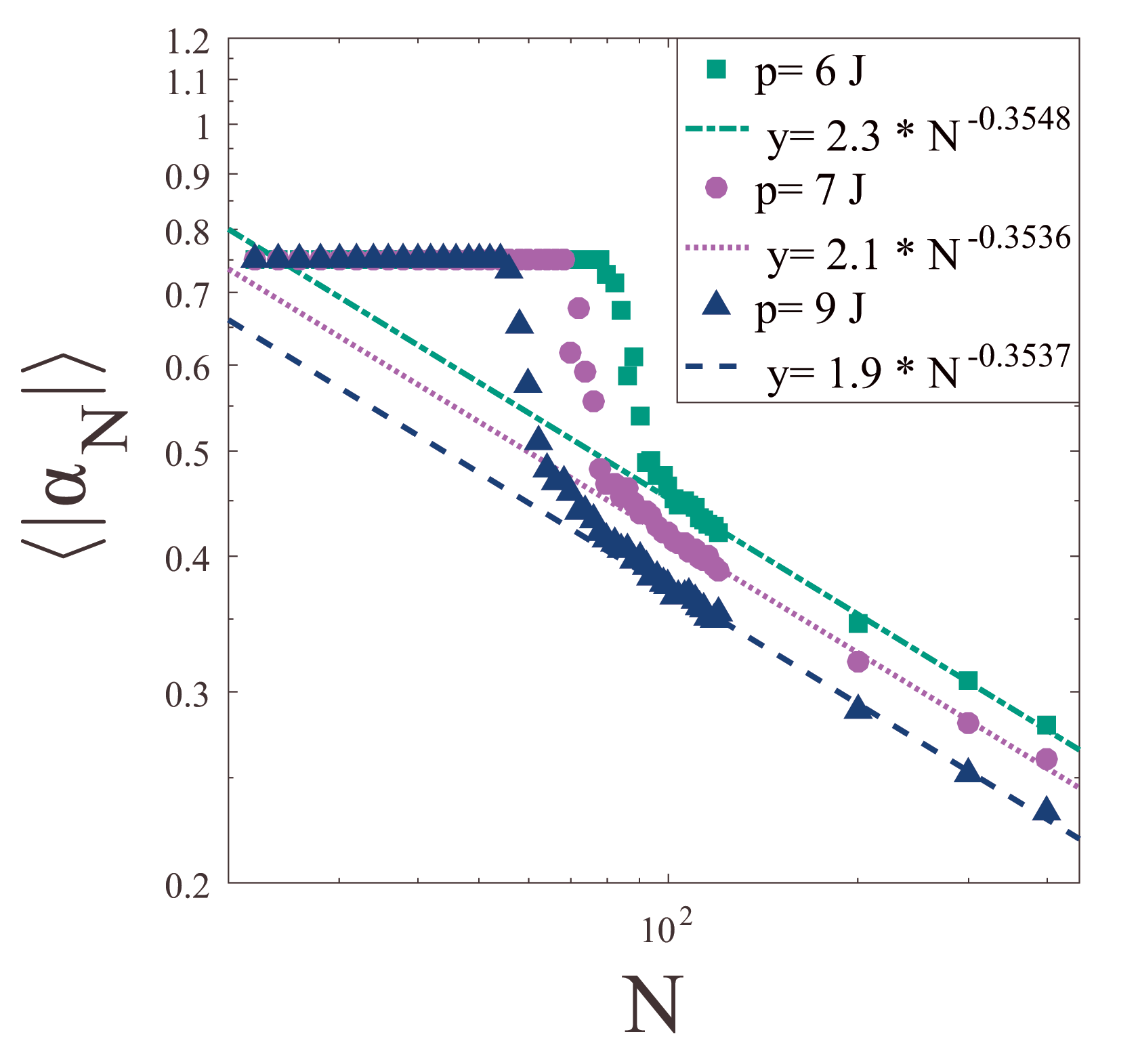}
\caption{Average transmitted field intensity $\langle|\alpha_N|\rangle$ as a function of chain length $N$ for different values of $p$ for fixed $U= 0.5J$. Markers denote the data points and their corresponding fits are shown with dotted lines.}
\label{F9}
\end{figure}

\begin{figure*}
\centering
\includegraphics[width=\textwidth]{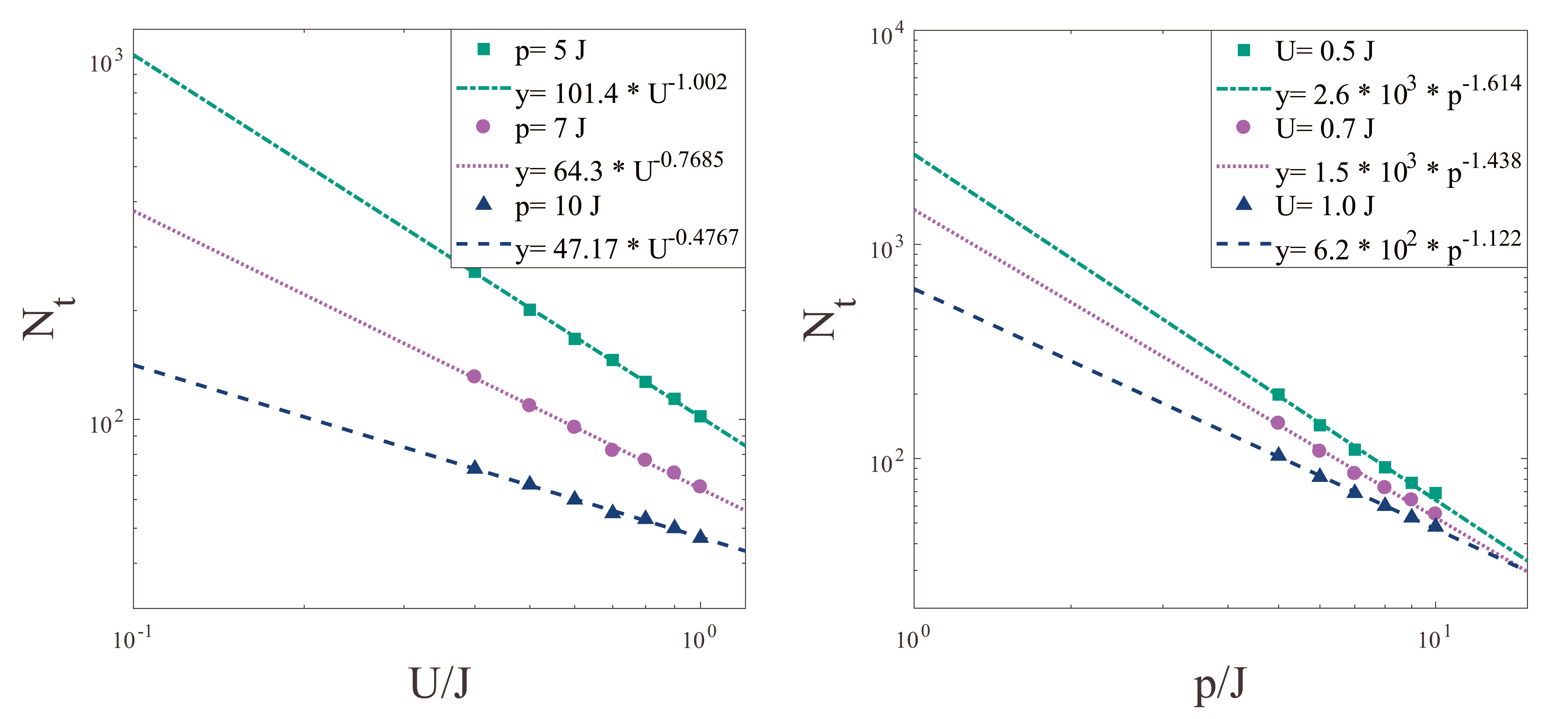}
\caption{(a) Threshold chain length $N_t$ as a function of $U$ for different values of $p$. (b) Threshold chain length $N_c$ as a function of $p$ for different values of $U$. Markers denote the data points and their corresponding fits are shown with dotted lines.}
\label{F1F2}
\end{figure*}

In what follows, the coupling constant $J$ will be used as an unit for all other model parameters, and will be assumed positive throughout. Very recently~\cite{vrecent} it has been shown that the physics of driven-dissipative Bose-Hubbard model remains invariant under change of signs of the system parameters. Positive values of $U$ are characteristic of the optical Kerr effect in resonant cavities, as discussed in~\cite{PhysRevB.85.033303}. While circuit QED settings are characterized by negative values of $U$ as shown in~\cite{Eichler2014} for a Bose-Hubbard dimer and in~\cite{PhysRevX.7.011016} for $72$ sites, it may be possible to conceive a system with positive $U$, for example in the dispersive limit of circuit QED~\cite{PhysRevA.79.013819} under strict choice of parameters. We believe that the Bose-Hubbard dimer setup of~\cite{Eichler2014} can be potentially scaled up to large system sizes to observe the effects described in this article.

Eqs.~(\ref{E2}-\ref{E4}) can be integrated using an adaptive fourth order Runge-Kutta method~\cite{Butcher:1996:HRM:230418.230420} with desired relative accuracy (in our case, $10^{-8}$). We focus our analysis on the average output field intensity $\langle|\alpha_N|\rangle$ for varying chain length $N$. As we will show below, the result displays an instability for a given range of the model parameters. For this reason, we average the field intensity over multiple realizations with randomized initial conditions. This instability is an intrinsic feature of the B-H model, due to two photon processes induced by the nonlinearity at resonant driving ($\omega_c=\omega_L$), and was already reported in analogous contexts \cite{PhysRevA.91.033823,PhysRevA.90.023827}. In the rest of this work, we will always assume zero detuning and set $\delta=0$ in all calculations.

\section{Results}\label{III}

To investigate the photonic transport along the chain, we begin by plotting the transmitted field intensity at the end boundary \textit{i.e.} $\langle|\alpha_N|\rangle$ for varying chain lengths $N$. Fig.~\ref{F0}(a) and (b) show the time-dependent intensity $\langle|\alpha_N|\rangle$ at the end site for $N=20$ and $N=100$ respectively, as computed for a given initial condition. While in the shorter chain a steady state is clearly reached, the longer chain shows a clear instability (which persists for integration times as long as computationally accessible). Even in the case of a long chain however, after a transient time a regime is reached where the time-averaged intensity is stationary. To better quantify this stationary state, we define an average value for the field intensity by averaging both over a time-window (well beyond the initial transient) and over an ensemble of randomized initial conditions.

Fig.~\ref{F1}(a) (log-log scale) displays the quantity $\langle|\alpha_N|\rangle$ as a function of chain length. For short chains, this value does not depend on the chain length, indicating ballistic transport. The ballistic regime coincides with the absence of instabilities, as illustrated in Fig.~\ref{F0}(a). Beyond a threshold chain length $N_t$, the transmission suddenly drops and the system starts exhibiting chaotic behavior as in Fig.~\ref{F0}(b). In this regime, the transmitted intensity $\langle|\alpha_N|\rangle|$ decays as a power law (with fitted curves displayed as dot-dashed lines), indicating a super-diffusive transport. The sudden onset of the instability is further highlighted by plotting the variance of the field intensity over time $\Sigma \equiv \sqrt{\langle|\alpha_N^2|\rangle - \langle|\alpha_N|\rangle^2}$ in Fig.~\ref{F1}(b) (semi-log scale). Zero variance denotes a steady state, while the sudden increase of the variance to a finite value marks the onset of the chaotic regime.

\begin{figure}
\centering
\includegraphics[width=0.5\textwidth]{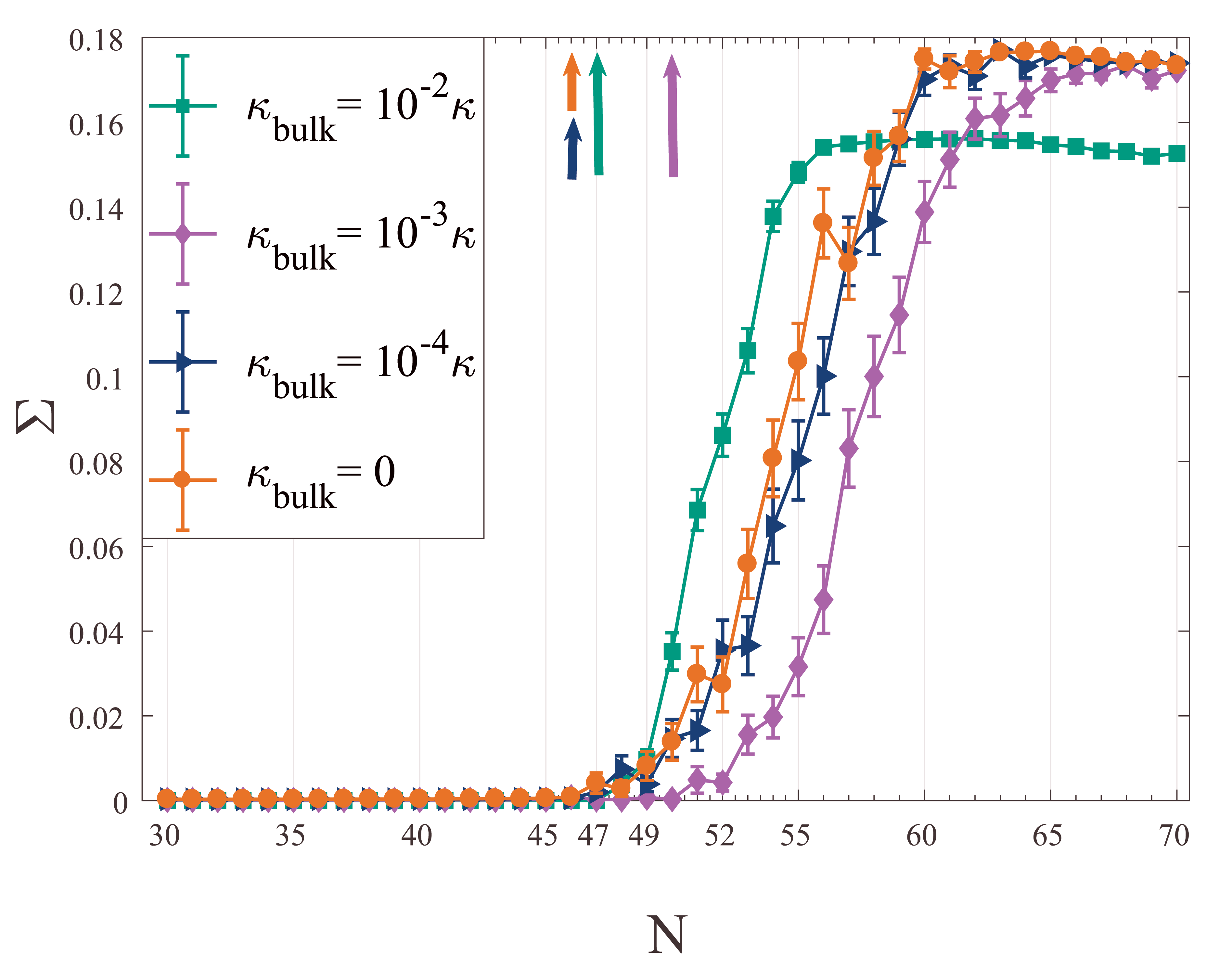}
\caption{The variance $\Sigma \equiv \sqrt{\langle|\alpha_N^2|\rangle - \langle|\alpha_N|\rangle^2}$ for small amount of bulk dissipation $\kappa_{bulk}$. The arrows in the figure indicate the point of origin of the chaos.}
\label{F00100}
\end{figure}

The mean field analysis thus predicts the onset of an instability beyond a threshold chain length $N_t$, marking a crossover between ballistic and super-diffusive photonic transport along the B-H chain. We define here $N_t$ as the chain length where the variance $\Sigma$ (averaged over realizations) increases beyond $0.05$. In Sec.~\ref{IV}, we shall discuss the onset of this instability from the perspective of a linearized stability analysis around stationary points very close to $N_t$. We argue here that the ballistic regime is a consequence of finite-size effects, while the transport properties in the thermodynamic limit are actually those corresponding to the super-diffusive regime found in the present analysis. The fact that the crossover between the two regimes as a function of $N$ is abrupt rather than gradual, can be ascribed to the strongly nonlinear character of the system under investigation. This feature plays an important role in view of the quest for dissipative phase transitions in photonic arrays. Indeed, if observed for a fixed chain length as a function of the system parameters $p$ or $U$, it may be inappropriately interpreted as the signature of an actual phase transition, while a correct analysis as a function of the chain length would bring to opposite conclusions. It is worth mentioning that the onset of chaotic behaviour in driven-dissipative Bose-Hubbard model has already been reported previously~\cite{PhysRevA.91.033823} in the context of soliton physics in a driven-dissipative Bose-Hubbard model. It was also observed in the experimental realization of~\cite{PhysRevX.7.011016}, where a stable regime resulted in a high power phase and the drop in transmission (low power phase) was followed by the onset of a chaotic regime, although the model was marginally different than the one discussed here.

In Fig.~\ref{F00100}, we show the variance $\Sigma$ for different values of bulk dissipation ($\kappa_{bulk}$) as a function of chain length. The variance $\Sigma$, clearly shows that the chaos survives even in the presence of a small bulk dissipation. We verified that this again gives rise to a diffusive transport as in the case of a non-dissipative bulk. For much higher values of bulk dissipation (\textit{$\kappa_{bulk}= \kappa$}), it is already known from previous studies ~\cite{Biella2015} that the photonic transport scales exponentially.

\begin{figure}
\centering
\includegraphics[width=0.48\textwidth]{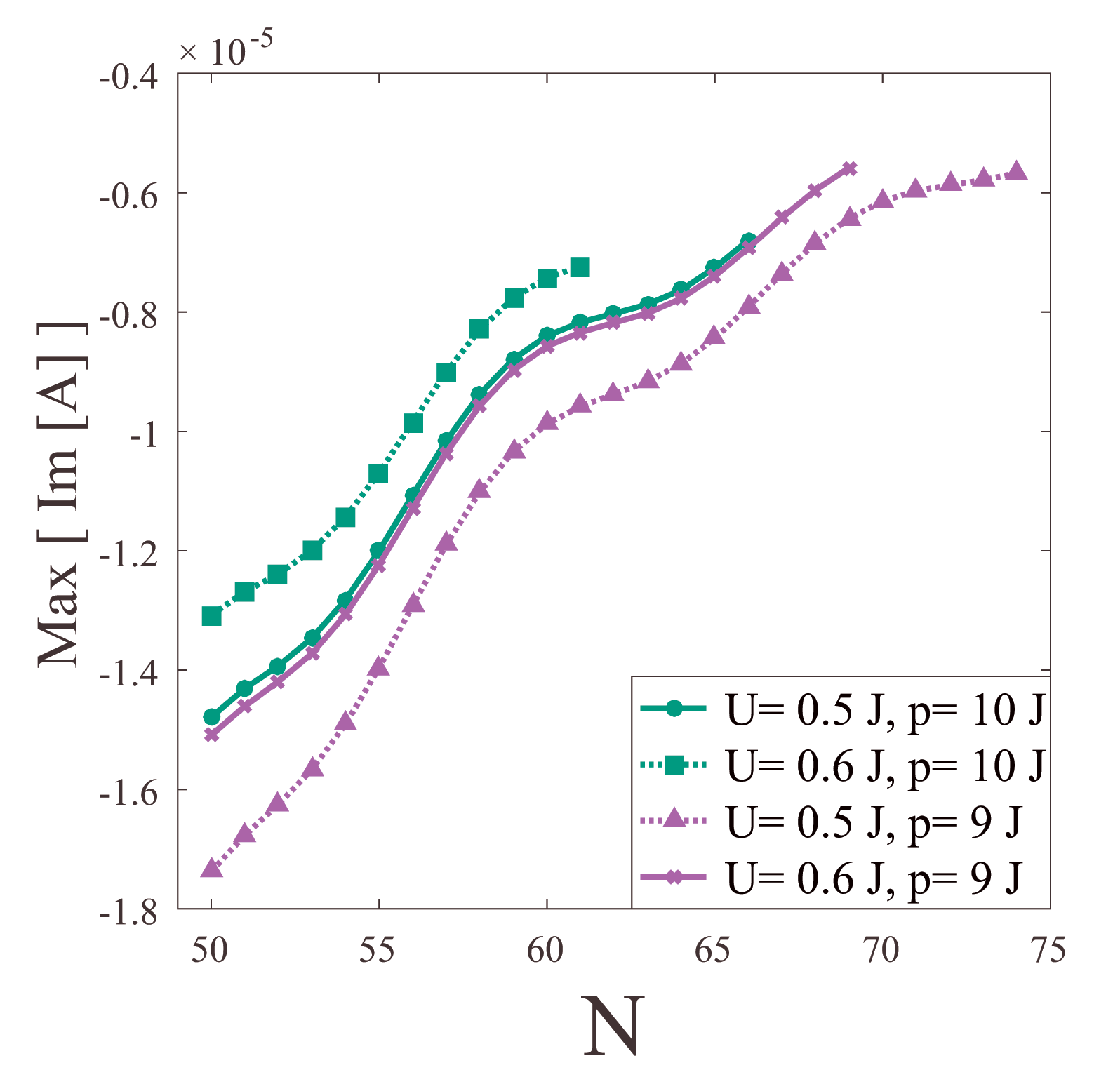}
\caption{Max [Im$\{E_\alpha\}$ ] as $N \rightarrow N_t$ for different values of $U$ and $p$. The stability analysis fails the moment $N$ touches threshold chain length $N_t$. A higher value of pump and nonlinearity can induce the instability much earlier.}
\label{F14}
\end{figure}

The threshold chain length $N_t$ depends on the intensity of the driving field, $p$ and on the nonlinear strength $U$, as illustrated in Fig.~\ref{F1} and Fig.~\ref{F9} respectively. A larger driving field (or nonlinear strength) can induce instability at smaller chain lengths. This can be understood by the fact that to a larger field correspond a larger total nonlinear energy associated to the Kerr medium, thereby limiting the range of finite-size effects to smaller chain lengths. Close to the threshold length $N_t$, the numerical simulations show that the onset of instability strongly depends on the choice of the initial condition, as studied in detail in the next Section.

To conclude this Section, we display in Fig.~\ref{F1F2}(a) and Fig.~\ref{F1F2} (b) the dependence of the threshold chain length $N_t$ on the nonlinear strength $U$ and on the driving field intensity $p$ respectively. In both cases we find a power-law dependence of $N_t$, which decreases as the two parameters are increased. The numerical simulations didn't provide any evidence of critical behavior at finite value of $U$ or $p$. Furthermore, the exponent characterizing the power law dependence on $U$ ($p$) is observed to vary continuously as a function of $p$ ($U$). This scenario corroborates the idea that transport in this model does not undergo any critical phenomenon. In the thermodynamic limit, ballistic transport persists only in two limits. In the limit of harmonic oscillators at zero nonlinearity the threshold chain length $N_t$ tends to infinity thus the ballistic scaling persists.
In the XX spin chain limit at infinite nonlinearity the power law exponent for the transmitted field tends to zero thus leaving a constant transmission regardless of the system size.
For any finite value of interaction the system always presents generalized diffusive transport with a power law scaling and finite chain length threshold $N_t$.
\begin{figure}[h!]
\centering
\includegraphics[width=0.48\textwidth]{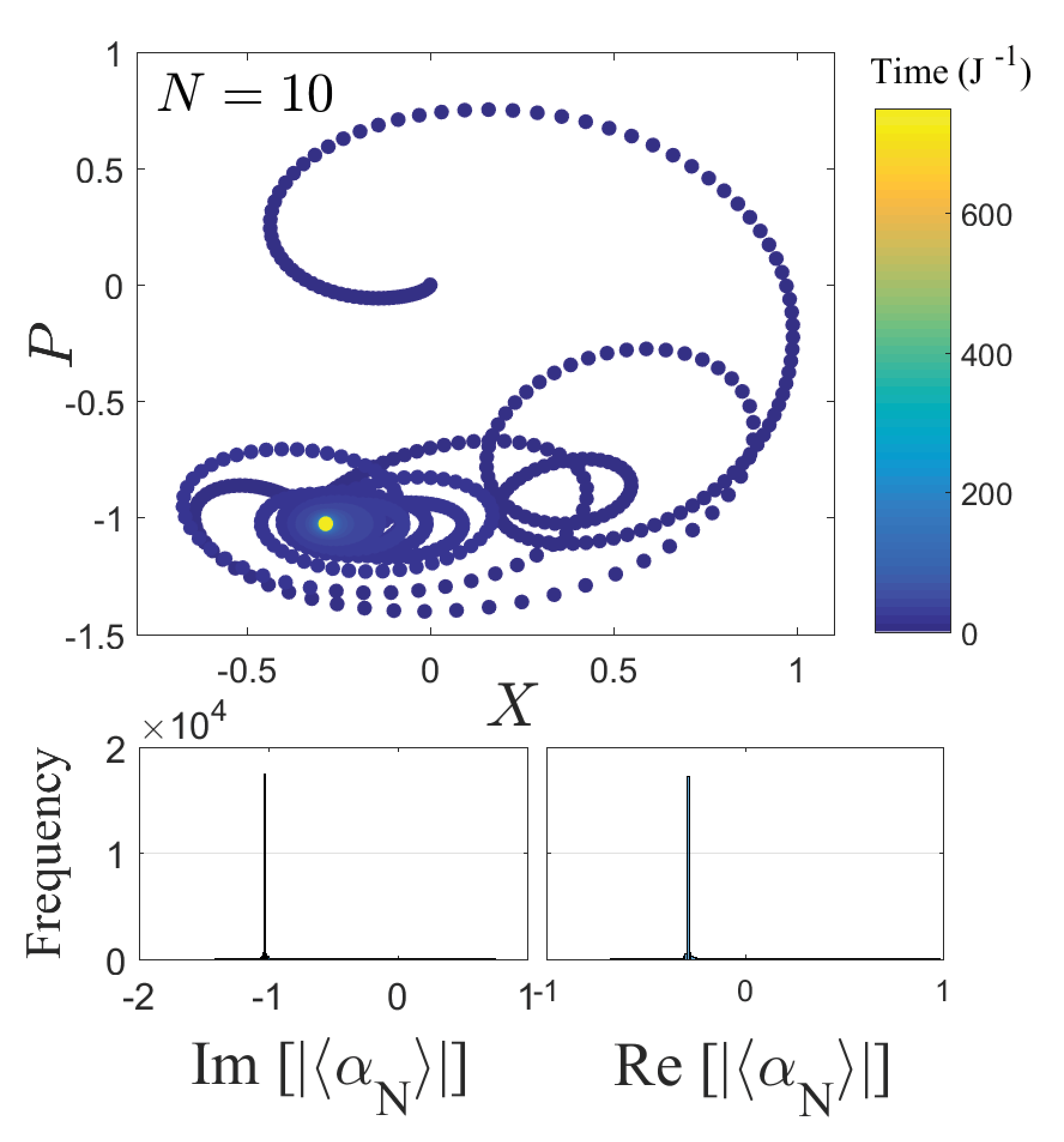}
\caption{Top panel- Time dynamics of $X \equiv (\langle \alpha_N \rangle + \langle \alpha_N^\dagger \rangle)$ and $P \equiv i(\langle \alpha_N \rangle -  \langle \alpha_N^\dagger \rangle)$ for $N= 10$, $U= 0.5J$ and $p= 10J$. Lower panel- Histogram of real and imaginary part of $|\langle\alpha_N\rangle|$, showing the frequency of occurrence.}
\label{F11}
\end{figure}
\begin{figure*}[h!]
\centering
\includegraphics[width=\textwidth]{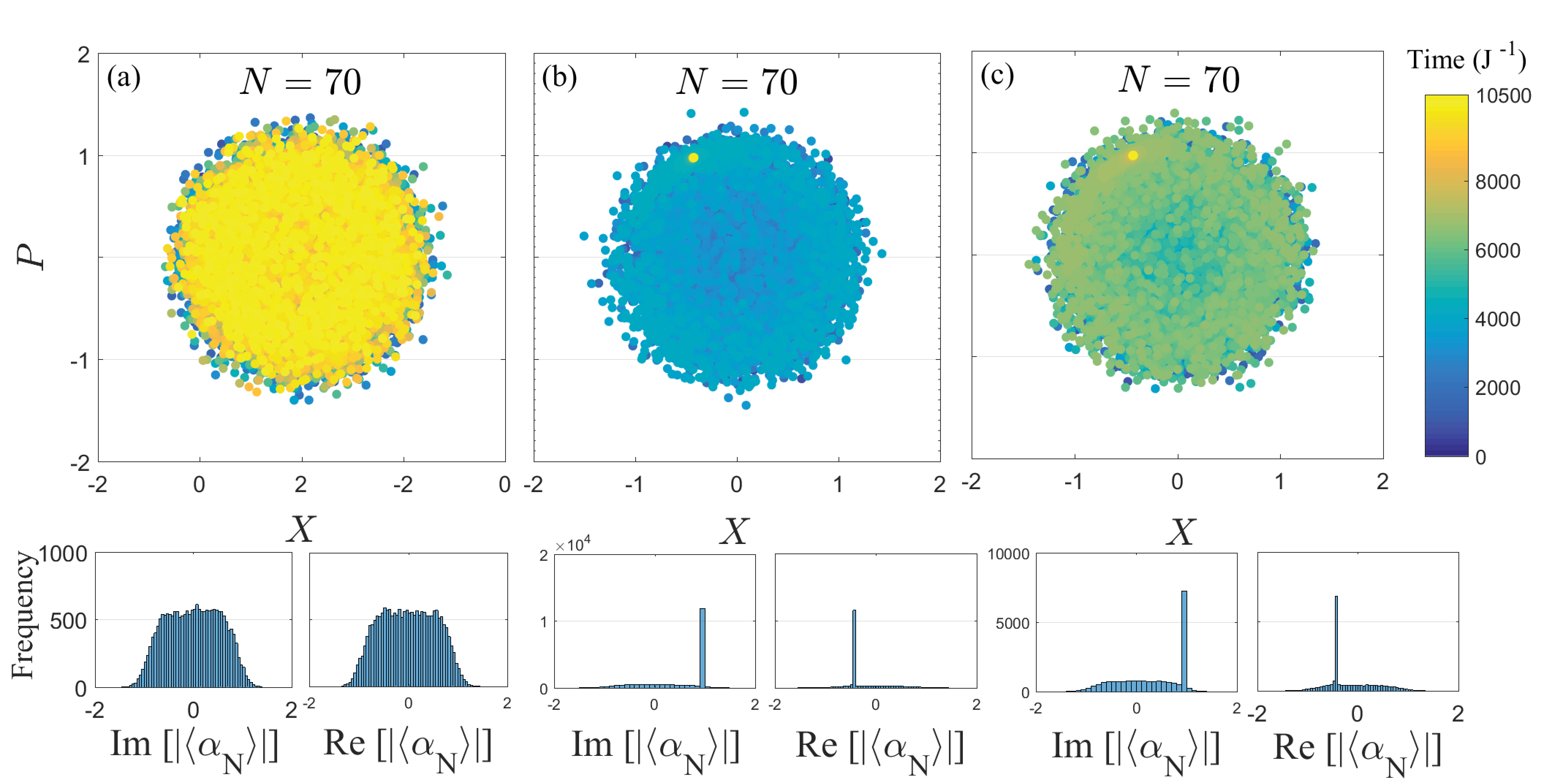}
\caption{Top panel- Time dynamics of $X \equiv (\langle \alpha_N \rangle +  \langle \alpha_N^\dagger \rangle)$ and $P \equiv i(\langle \alpha_N \rangle -  \langle \alpha_N^\dagger \rangle)$ for $N= 70$, $U= 0.5J$ and $p= 10J$ with (a) zero  (b), (c) different random initial conditions. Lower panel- Histogram of real and imaginary part of $|\langle\alpha_N\rangle|$, showing the frequency of occurrence.}
\label{F12}
\end{figure*}
\begin{figure*}[h!]
\centering
\includegraphics[width=0.48\textwidth]{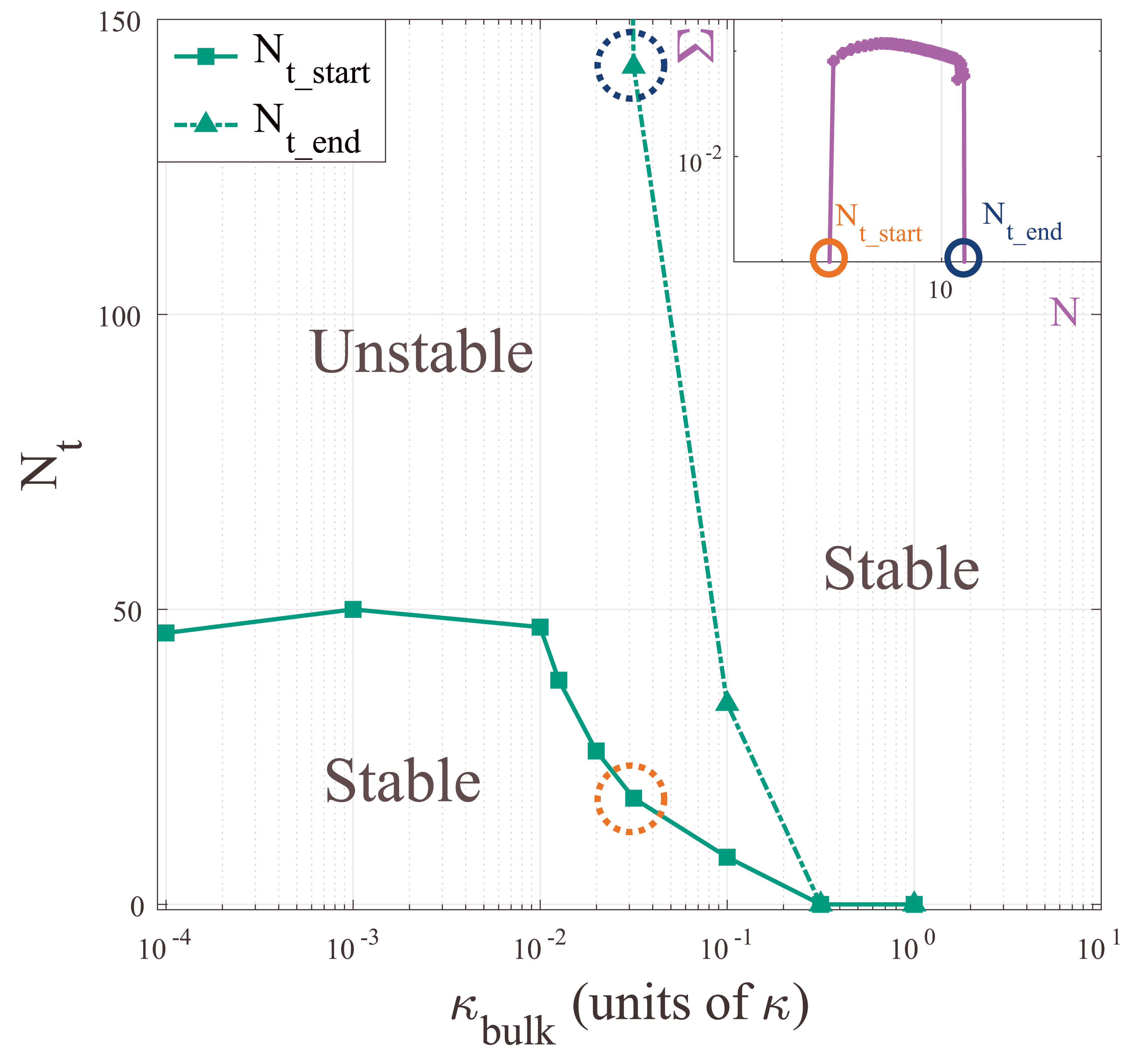}
\caption{$N_t$ as a function of $\kappa_{bulk}$ highlighting the stable and unstable regime. Inset- The variance $\Sigma$ showing the onset and collapse of chaotic regime at $N_{t\_start}$ (red circle) and $N_{t\_end}$ (blue circle) respectively.}
\label{forgot}
\end{figure*}

 As a final remark, we notice that the power-law scaling displayed in Figs.~\ref{F1F2}(a) and (b) allow to infer the behavior of the system for smaller values of the nonlinearity $U$ and/or the driving field amplitude $p$, which may be more appropriate to existing experimental platforms in the domain of superconducting circuits. A direct numerical study of these regimes on the other hand may be challenging, as the threshold behavior would appear at much longer chaing lengths.

\subsection{Effect of $\kappa_{bulk}$}
As mentioned previously, the bulk dissipation plays a major role in determining the stability of the system. It is clear from the results presented in this article that the photonic transport scales as power law as a function of chain length when one completely neglects the dissipation of the bulk. For the extreme case, when the bulk dissipation becomes equal to the boundary dissipation (\textit{i.e.} $\kappa_{bulk}= \kappa$), it is already known~\cite{PhysRevA.91.053815} that the system is stable and the photonic transport scales exponentially. It is therefore crucial to investigate how the threshold chain length $N_t$ varies between these two limits. In Fig.~\ref{forgot}, we plot $N_t$ as a function of bulk dissipation. As $\kappa_{bulk}$ is increased, the threshold chain length $N_t$ (square markers) decreases and eventually becomes zero when $\kappa_{bulk}$ approaches $\kappa$. Interestingly, just before that regime, a small region emerges where the onset of chaos is followed by its collapse at longer chain lengths (we called it $N_{t\_end}$ and is shown with triangular markers). This is evidenced by the variance $\Sigma$ as a function of chain length (inset of Fig.~\ref{forgot}). This behavior can be interpreted as follows. For any small but finite value of $\kappa_{bulk}$, there must be a sufficiently long chain length such that the combined effect of dissipation on each site is enough to overcome the nonlinearity and make the system stable. At the same time, for short enough chains, a small enough value of $\kappa_{bulk}$ should produce a marginal effect and the system is essentially expected to behave as for the $\kappa_{bulk}=0$ case. As a result, for small enough $\kappa_{bulk}$, chaos arises for intermediate chain length, while very short and very long chains will be stable. This behavior is illustrated in Fig.~\ref{forgot}.

\section{Stability Analysis}\label{IV}
To gain further insight into the mechanism underlying the onset of instability, we perform a stability analysis by studying the eigen-energy of the linearized excitations of the stable solution. We can only perform this analysis for parameters values where a steady-state solution is numerically accessible, namely in the stable region. In spite of this limitation however, we can still extract useful information from this analysis by studying the behavior of the excitations when approaching the onset of the instability. The steady state solution can be obtained by setting the time derivative to zero in Eqs.~(\ref{E2}-\ref{E4}), and solving the corresponding algebraic equations for $\alpha^{(s)}_j$. Following the standard approach \cite{PhysRevB.77.125324}, we then assume that the excitations of this solution take the form
\be
\alpha_j(t)= e^{-i\omega_L t}[\alpha^{(s)}_j + \tilde{\delta} \alpha_j(t)]\,,
\label{E5}
\ee
around the stationary points $\alpha^{(s)}_j$, where $\omega_L$ is the frequency of the driving field on the first site. Substituting the above ansatz in Eqs.~(\ref{E2}-\ref{E4}) and neglecting terms that are nonlinear in $ \tilde{\delta} \alpha_j(t)$, one obtains
\be
\ba
\dot{\alpha_j}= &-&i\frac{\kappa_j}{2} \tilde{\delta}\alpha_j \delta_{j,(1,N)} + 2U|\alpha^{(s)}_j|^2\tilde{\delta}\alpha_j + U(\alpha^{(s)}_j)^2\tilde{\delta}\alpha^*_j \\
&-& J\tilde{\delta}\alpha_{j+1} \delta_{j,(1,2...N-1)} - J\tilde{\delta}\alpha_{j-1} \delta_{j,(2...N)}
\label{E6}
\ea
\ee
$\delta_{j,(1,N)}$ is a Kronecker-delta function accounting for the drive and dissipation on the first and last sites. Introducing the Bogoliubov transformation
\be
\tilde{\delta}\alpha_j (t)= e^{-iEt/\hbar} U_j + e^{iE^*t/\hbar}V^*_j
\label{E7}
\ee
and substituting it in Eq.~\ref{E6}, the problem reduces to a secular equation of type
\be
A\tilde{\delta}\Psi= E \tilde{\delta}\Psi
\ee
where $\tilde{\delta}\Psi = (U_1 V_1 U_2 V_2 ....U_N V_N)^T$ and the matrix $A$ is expressed as
$$
\begin{pmatrix}
-i\frac{\kappa_1}{2} + 2Un_1 & U\alpha^{s2}_1 & -J & 0 &0 &0 &\cdots\\
-U(\alpha^{s*}_1)^2 & -i\frac{\kappa_1}{2} - 2Un_1 &0 & J &0 &0 &\cdots\\
-J  & 0  & 2Un_2  & U\alpha^{s2}_2 & -J & 0  &\cdots\\
0 & J  & -U\alpha^{s2}_2  & -2Un_2 & 0& J&\cdots \\
\vdots  &\vdots  &\vdots &\vdots &\vdots &\vdots&\ddots
\end{pmatrix}
$$
where $n_j= |\alpha^{(s)}_j|^2$. An instability in the eigenvalues of this secular problem is signaled by the imaginary part of any of the eigen-frequencies becoming larger than zero. In Fig.~\ref{F14}, we plot the maximum of Im$\{E_\alpha\}$ as computed for various system parameters and for increasing chain length $N$ within the stable region. A trend is clearly visible, whereby this quantity approaches zero and eventually positive values as $N$ increases. The actual instability points could not be reached because of the numerical difficulty in finding the steady state solution. Indeed, as discussed below, the onset of chaos is preceded by solutions displaying limit cycles or multi-stability. In spite of these limitations, the result shows rather clearly that the onset of chaos is triggered by unstable excitations enforced by the increasingly strong nonlinearity that is associated to the increasing chain length.

Finally, we study the time evolution of the field at the last chain site $\langle\alpha_{N}(t)\rangle$. Fig.~\ref{F11} displays the result for $N=10$, for which steady state is expected. The color bar denotes the time $t$ in units of $J^{-1}$. The steady state is reached at around $Jt= 200$. The histogram in the lower panel shows the frequency of occurrence of imaginary ($P$) and real ($X$) values of $\langle\alpha_N\rangle$, highlighting the occurrence of the steady state. A similar analysis is performed for $N=70$, namely close to the threshold chain length, and the result is shown in Fig.~\ref{F12}. The three plots correspond to three different choices of the initial condition. Panel (a) in particular shows the result for zero initial condition. Here no steady state is reached and the system evolves chaotically along a trajectory which densely covers a bound region of phase space, as also shown by the corresponding histograms. For such values of $N\sim N_t$ we notice that, for different initial conditions, the system has finite likelihood to undergo a chaotic trajectory or to achieve a steady state. Fig.~\ref{F12}(b) and (c) display two cases of the latter type. In panel (b) the steady state is reached more rapidly, while for the case in panel (c) a metastable strange attractor appears (the faint ring shape in the plot) and the steady state is reached at a later time. A qualitative analysis of these three cases indicates that, when approaching the threshold length from below, an increasing number of long-lived metastable patterns characterizes the trajectories in phase space, with a concomitantly larger dependence on the initial conditions, until eventually the steady state leaves room to a completely chaotic behavior.

\section{Disorder and metal-insulator transition}\label{V}
\begin{figure}
\centering
\includegraphics[width=0.5\textwidth]{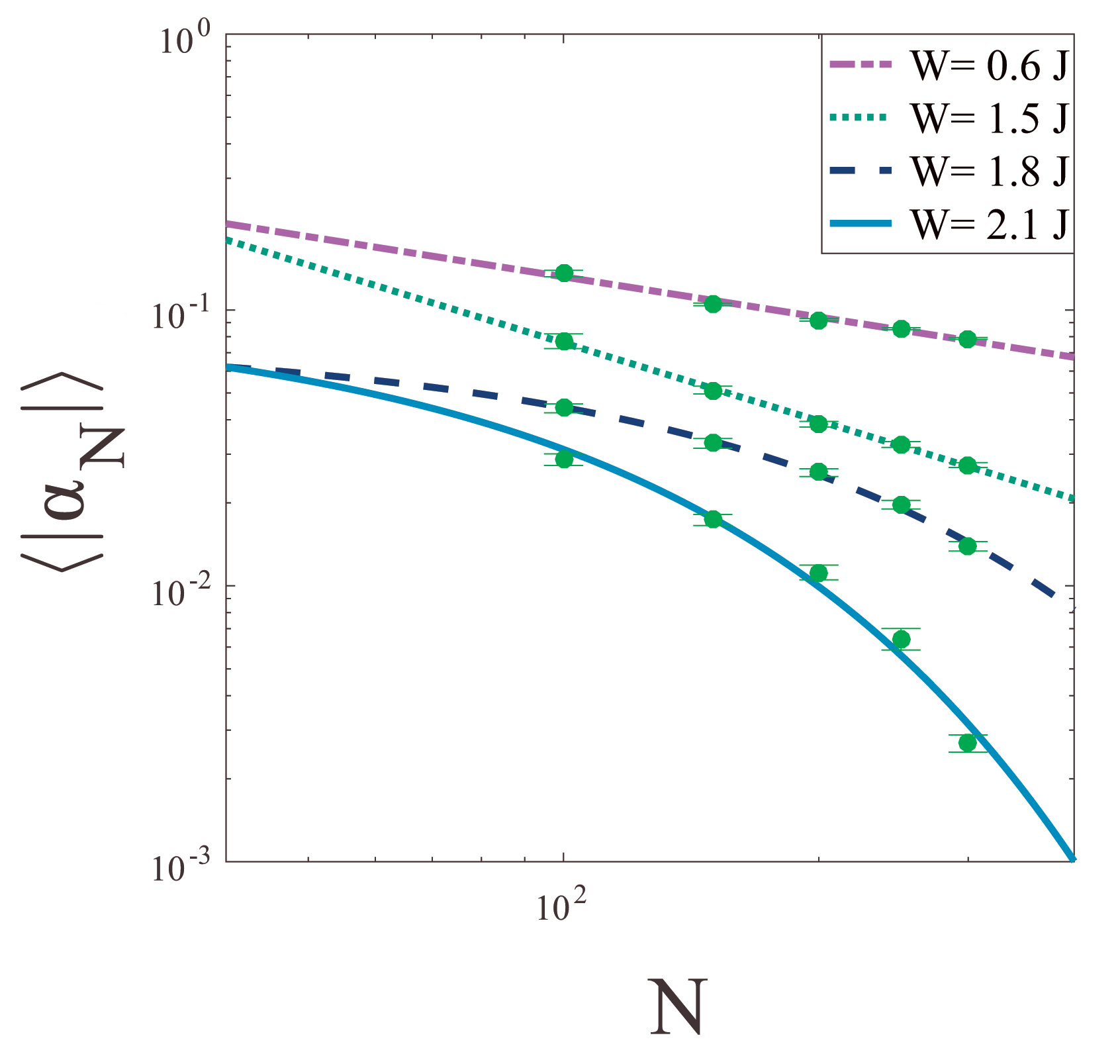}
\caption{$\langle|\alpha_N|\rangle$ for $N= 100,200,300,400,500$ for different values of disorder $W$. Parameters used are $U= 5J$, $p= 10J$, $\kappa= J$ and $\delta=0$.}
\label{F15}
\end{figure}
The finding that optical transport is diffusive raises a natural question, namely whether disorder may induce a transition to an insulating regime \cite{PhysRevA.90.031603}. Here, we extend the numerical study to the case in presence of disorder and study the scaling of the transmitted field intensity with the chain length, as done above. We model disorder as a random fluctuation of the local frequency of the oscillators $\tilde{\omega}_j= \omega_c +\xi_j$, where $\xi_j$ is uniformly distributed in the interval $[-W,W]$. The analysis that follows was carried out by performing a configuration average over 100-150 disorder realizations, for each set of chain parameters under study. Fig.~\ref{F15} displays the output field intensity as a function of chain length for different values of the disorder parameter $W$. On a log-log scale, within the range of chain length considered, the result displays a rather well defined transition from power-law to exponential decay as $W$ is increased.
\begin{figure}
\centering
\includegraphics[width=0.45\textwidth]{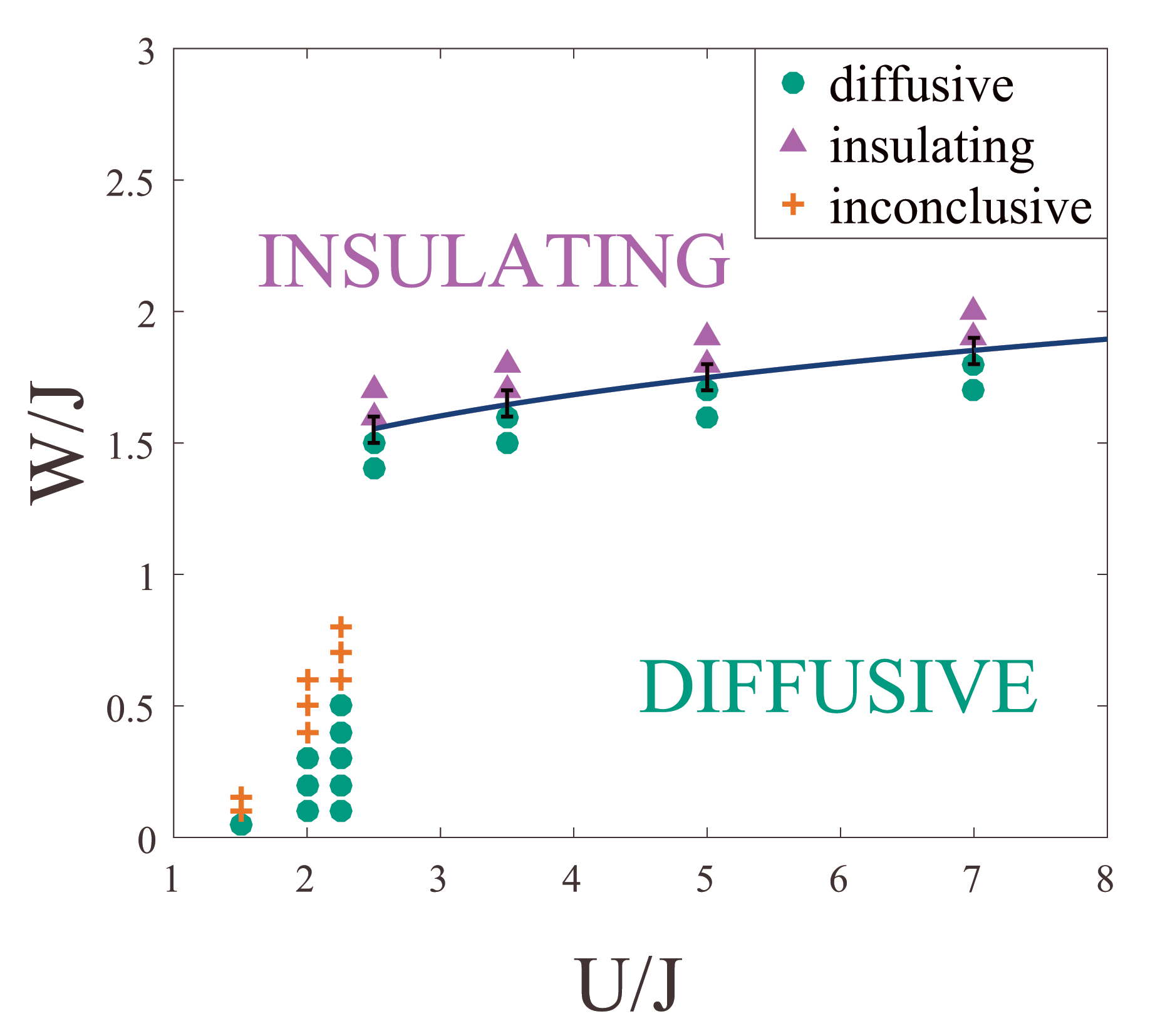}
\caption{The phase diagram as a function of nonlinearity and disorder. Inconclusive results keeps the phase boundary indeterministic in small $U$ regime.}
\label{F16}
\end{figure}
In Fig.~\ref{F16} we present the summary of the results obtained for various values of disorder and nonlinearity. For some of the points, the numerical simulation didn't allow to discriminate between the diffusive and insulating behavior, due to lack of numerical accuracy for the largest values of the chain length that were required. Still, the data highlight a part of the phase boundary at values of $U/J>2$.

\section{Conclusion}\label{VI}
To summarize, we have theoretically investigated the emergent physics and transport behaviour in a boundary driven Bose Hubbard chain. One important result is the sudden onset of instability at a threshold chain length $N_t$, followed by an abrupt drop in transmission as the signature of generalized diffusive transport. We showed that this threshold chain length depends on the intensity of the pump and nonlinearity. This reveals that for a single sample system, increasing the intensity of the pump would induce a drop of the transmission which should not be interpreted as a standard phase transition but rather as the regime in which finite size effects give way to emergent phenomena. In the thermodynamic limit and in the absence of disorder, the system always displays generalized diffusive transport, which becomes ballistic only in the limits of zero and infinite nonlinearity \cite{PhysRevE.87.012109,PhysRevLett.117.040601}. Finally, we have shown the competition between disorder and nonlinearity and how it induces a diffusive-insulator phase transition.

 A natural question arises about how the chaotic behavior, emerging from the present classical field model, will be described in a quantum treatment of the driven-dissipative system, for which a unique steady state should still be expected for a system of finite size. To this purpose, while solving a pure quantum model using the master equation may be a formidable task in terms of computational resources, approximate techniques like phase space methods may provide insight into the quantum fluctuations how they relate to the classical chaos observed here.

As a final remark, it will be interesting to establish a closer link between the vanishing of transport in disordered bosonic systems and the field of many-body localization which has mostly focused on fermion and spin systems~\cite{PhysRevLett.117.040601,Kd1234, PhysRevB.91.081103,PhysRevLett.114.160401,PhysRevB.82.174411}.

\bibliography{Reference1}
\end{document}